# Semantics-aware Virtual Machine Image Management in IaaS Clouds

Nishant Saurabh*†, Julian Remmers†, Dragi Kimovski*, Radu Prodan*† and Jorge G. Barbosa‡
*Institute of Information Technology, University of Klagenfurt, Austria
†Institute of Computer Science, University of Innsbruck, Austria
‡LIACC, Faculdade de Engenharia da Universidade do Porto, Portugal

*Abstract*—Infrastructure-as-a-service (IaaS) Clouds concurrently accommodate diverse sets of user requests, requiring an efficient strategy for storing and retrieving virtual machine images (VMIs) at a large scale. The VMI storage management require dealing with multiple VMIs, typically in the magnitude of gigabytes, which entails VMI sprawl issues hindering the elastic resource management and provisioning. Nevertheless, existing techniques to facilitate VMI management overlook VMI semantics (i.e at the level of base image and software packages) with either restricted possibility to identify and extract reusable functionalities or with higher VMI publish and retrieval overheads. In this paper, we design, implement and evaluate Expelliarmus, a novel VMI management system that helps to minimize storage, publish and retrieval overheads. To achieve this goal, Expelliarmus incorporates three complementary features. First, it makes use of VMIs modelled as semantic graphs to expedite the similarity computation between multiple VMIs. Second, Expelliarmus provides a semantic aware VMI decomposition and base image selection to extract and store non-redundant base image and software packages. Third, Expelliarmus can also assemble VMIs based on the required software packages upon user request. We evaluate Expelliarmus through a representative set of synthetic Cloud VMIs on the real test-bed. Experimental results show that our semantic-centric approach is able to optimize repository size by $2.2 - 16$ times compared to state-of-the-art systems (e.g. IBM's Mirage and Hemera) with significant VMI publish and retrieval performance improvement.

*Keywords*—Virtual machine image management, semantic similarity, storage optimization.

## I. INTRODUCTION

The evolving Cloud architecture [4], [7], [17], [25], [28] requires efficient and scalable on-demand provisioning and management of computing services over a federated and heterogeneous infrastructures. Virtualization [2], [24] emerged as a key technology for enabling and provisioning of such computing services. One common virtualization technique that facilitates the computing services is the virtual machine (VM) [6], [8], [26], [27], instantiated using a user-created template called VM image (VMI) [13]. Such VMIs comprise an operating system (OS) and user-specific customized software package(s). The ever increasing number with size of each VMI in the magnitude of gigabytes induces important management issues such as VMI sprawl [23], hindering the elastic resource management and provisioning process. For example, Amazon Elastic Compute Cloud (EC2) alone consists of more than $30,000$ public VMIs [3], where typical operations like cloning, versioning, sharing, storing and transforming VMIs in dedicated repositories introduce a high amount of storage redundancy and maintenance costs.

To solve VMI management challenges such as sprawl, prior research in this domain primarily focused on leveraging VMI deduplication [14], [16], [18] and caching [11], [19], [22], [29] by identifying similar byte segments [10], [12], [20]. Such techniques optimize the VMI storage and reduce redundant content by up to 80%, but limit the benefits of the virtualization technology, such as exploiting stronger isolation between software packages, and keeping a provenance record of changes and reusable functionality in the VMI at a semantic level [23].

Nevertheless, few of the recent works, namely IBM's Mirage [23] and Hemera [15] improve upon the previous studies and explore VMI management at a more fine grained file-system level. Both systems improve upon the VMI sprawl, but with significant VMI publishing and retrieval overheads.

To address these challenges, we propose a novel VMI management system called Expelliarmus that represents the VMI and its components as structured graph [1], presenting the functional requirements between the base image and different software packages within the VMI. For each VMI semantic graph, we also extract two induced subgraphs, called the base image subgraph and software package subgraph. The purpose of this operation is to merge one or more semantically similar VMIs into a single VMI master graph, clustering the software packages of one or more VMIs with a semantically similar base image. This approach reduces the similarity computation overhead by comparing every VMI to the master graph instead of individual VMI graphs. To facilitate the semantic aware decomposition, we employ techniques to only extract the unique software packages that are not yet existing in the repository. Moreover, we devise a base image selection algorithm and define a novel semantic compatibility metric to select from a pool of semantically similar base images, one that is functionally compatible to existing software packages replacing the redundant ones. Our approach also provide means for VMI assembly, either with identical uploaded software packages or compatible with a base image already existing in the repository. To study the benefits of semantics aware VMI management, we performed an extensive series of experiments using a representative set of synthetic Cloud images, currently limited to `Linux` VMIs only. The results demonstrate that Expelliarmus significantly optimizes the storage cost with improved VMI publish and retrieval times compared to related [15], [23] state-of-the-art systems.

The paper is organized as follows. Section II summarizes the related work. Section III presents the model together with the VMI semantic representation, and formulates the

VMI semantic similarity and compatibility metrics. Section IV describes the architecture of the Expelliarmus system and its design components, including the VMI publishing, base image selection and VMI retrieval algorithms. Section V provides implementation details and Section VI presents the experimental results. Finally, Section VII concludes the paper.

## II. RELATED WORK

One key factor affecting the performance of IaaS Cloud management systems is the rapidly increasing number and size of each stored VMI exceeding multiple GB. This introduces critical challenges to VMI management such as VMI sprawl, and hence impacts elastic resource management and provisioning. Moreover, these VMIs are usually excessively similar with a high a high degree of redundancy, addressed in the community through deduplication, due to its wide adaptation in archiving systems.

Jin et al. [11] explored the effectiveness of the block level deduplication, both with fixed and variable size chunking using Rabin fingerprinting [21] schemes. They showed that VMIs with the same guest OS and different software packages share considerable amount of data. This study also emphasizes that VMI deduplication at block level with fixed size chunking scheme is more efficient than variable size chunking, detecting up to 70% of identical content between VMIs.

Jayram et al. [10] built upon similar work comparing different VMI deduplication techniques and providing metrics for estimating VMI similarity. Their study showed that the appropriate chunk size selection is essential to decide the block level deduplication factor used for similarity computation.

Zhao et al. [31] proposed a scalable VMI file system called Liquid that enables large scale VM deployment through a fixed size block level deduplication, resulting in a low storage consumption. The system also improves the I/O performance with a peer-to-peer networked VMI sharing and distribution.

Chun-Ho et al. [18] took a step forward and propose a VMI backup system based on a block level reverse deduplication that removes duplicates from old VMIs, while keeping the new VMI layout as sequential as possible.

Xu et al. [29], [30] proposed a VMI backup system named Crab that also uses deduplication at block level, but implements an additional a k-means clustering method to group VMIs and consequently speedup the index lookup overhead.

Reimer et al. [23] and Ammons et al. [1] proposed a new VMI format called Mirage represented as VMI structured data, performing file system indexing together with file level deduplication to improve the inventory control and VM deployment.

Liu et al. [15] also approached the VMI as structured data in a rigorous database structure called Hemera that, in contrast to the Mirage, transforms the VMI operations into database operations based on simple SQL queries.

All these works proposed optimization to VMI management and employed VMI deduplication at content level with restricted possibility to identify and extract reusable functionalities. Although, Mirage and Hemera use additional semantic data to access VMI at a more fine grained file-system level, but with significant VMI publishing and retrieval overheads.

Our work improve over these approaches on both aspects. Instead of splitting the VMI into chunks or files, our semantic-aware decomposition and base image selection approach splits the VMI at semantic level into a base image and one or more software packages, such that no redundant package and base image is stored twice. On the contrary, aforementioned approaches store additional non-redundant content of all base images. Moreover, we assemble the VMIs not necessarily with the same base image, but also using a semantically similar one. Second, semantic-centric optimization to VMI storage exploited by our system, reduces the VMI content size to transfer from and into the repository with significant VMI publish and retrieval performance improvement.

## III. VMI SEMANTIC MODEL

This section presents a formal model and a set of basic definitions essential to this work.

### A. Virtual machine image (VMI)

A *virtual machine image (VMI)* $I = (BI, PS, DS, Data)$ consists of a *base image* $BI$ with a standalone OS, a set of software packages $PS$ and $DS$ installed on top, ranging from database to application servers, and a set user data $Data$.

A *primary package* set $PS$ is a suite of software packages eligible to be hosted on an OS within a VMI. We assume that every VMI consists of one or more primary packages, required by the user upon instantiation.

A *dependency package* set $DS$ contains libraries or other packages internal or external to the OS within the base image, used to build or install the primary packages within the VMI.

The $Data$ component corresponds to the *user data* (e.g. files, directories) not recognized by the guest OS package management (e.g. `home` directory in a Linux file system).

### B. VMI semantic graph

The *VMI semantic graph* is a high-level intermediate graph representation expressing the rich VMI structure in terms of functional requirements and relationships between base image, primary packages, and dependency packages. We define the semantic graph of a VMI $I$ as a directed cyclic graph $G_I = (V_I, E_I)$, where $V_I = BI \cup PS \cup DS$ represents the set of vertices including the base image, primary and dependency packages, and $E_I \subseteq V_I \times V_I$ is the set of edges, where a direct edge $e = (v, v') \in E_I$ denotes a dependency of the base image, primary package, or dependency package $v$ on $v'$.

Figure 1a shows a typical VMI semantic graph consisting of a `Debian` base image, two primary packages (`MariaDB` and `Tomcat8`), and several required dependency packages (i.e. `bash`, `openjdk`, `gawk`, `libc6`, `dpkg`, `debconf`, `perl-base`, `ucf` and `coreutils`). The `libc`, `perl-base` and `dpkg` have a cyclic dependency, meaning that always they need to be provided and installed together.

### C. VMI attributes

Each base image $BI$ of a VMI $I$ has a quadruple of *attributes* $attrs(BI) = (type, distro, ver, arch)$, expressing the guest OS name or $type$ (e.g. `Linux`), its distribution

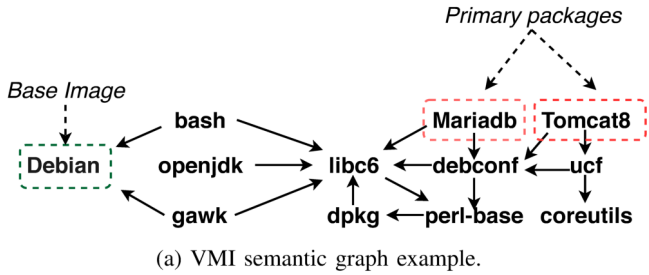

(a) VMI semantic graph example.

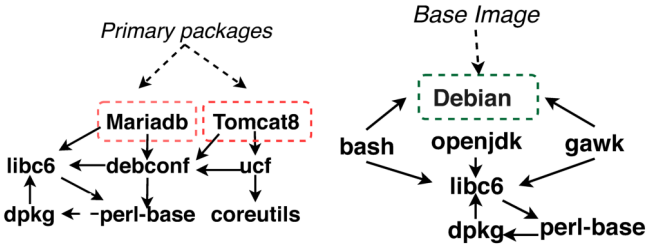

(b) VMI primary package subgraph for the VMI semantic graph from Figure 1a.

(c) Base image subgraph for the VMI semantic graph from Figure 1a.

Fig. 1: VMI semantic graph, primary image subgraph, and base image subgraph example.

TABLE I: Attributes of a VMI semantic graph $G_I = (BI, PS, DS, Data)$.

| Semantic attribute | Example |
|---|---|
| $type : BI \mapsto String$ | $type(BI)$ = "Linux" |
| $distro : BI \mapsto String$ | $distro(BI)$ = "debian" |
| $ver : BI \mapsto String$ | $ver(BI)$ = "16.04" |
| $arch : BI \mapsto String$ | $arch(I)$ = "x86_64" |
| $pkg : PS \cup DS \mapsto String$ | $pkg(P)$ = "Tomcat8" |
| $ver : PS \cup DS \mapsto String$ | $ver(P)$ = "8.1.1" |
| $arch : PS \cup DS \mapsto String$ | $arch(P)$ = "multiarch" |
| $size : PS \cup DS \mapsto String$ | $size(P)$ = "4096" |

$distro$ (e.g. Debian), its version $ver$ (e.g. 16.04), and its architecture $arch$ (e.g. x86_64).

Similarly, each primary or dependency package $P \in PS \cup DS$ holds a quadruple of attribute information $attrs(P) = (pkg, ver, arch, size)$ about its name $pkg$ (e.g. MariaDB), version $ver$ (e.g. 2.5), architecture $arch$ (e.g. multiarch), and $size$ (e.g. 1056).

Further, we prohibit a VMI semantic graph $G_I = (V_I, E_I)$ to contain multiple packages with the same $pkg$ attribute, i.e. $\forall (P_1, P_2) \in V_I, pkg(P_1) \neq pkg(P_2)$. Table I summarizes the attributes associated to a VMI and its semantic graph.

### D. Semantic graph union

We define the *union of two semantic graphs* $G_1 = (V_1, E_1)$ and $G_2 = (V_2, E_2)$ as a new semantic graph $G = G_1 \cup G_2$, where $G = (V_1 \cup V_2, E_1 \cup E_2)$. Two vertices $v_1 \in V_1$ and $v_2 \in V_2$ are equivalent and belong to the intersection of the vertice sets of the two semantic graphs ($v_1 \equiv v_2 \in V_1 \cap V_2$) if they have the same attribute values: $attrs(v_1) = attrs(v_2)$.

### E. VMI subgraphs

A *primary package subgraph* $G_I[PS]$ is an induced subgraph of the VMI semantic graph containing all primary packages $PS \subset V_I$ and dependency packages reachable from them as vertices:

$$G_I[PS] = \bigcup_{\forall P \in PS} G_I[P],$$

where $G_I[P]$ is the induced connected subgraph containing all packages reachable from a package $P$:

$$G_I[P] = (V_P, E_P), P \in V_P \subset I \wedge E_P \subset E \wedge$$
$$\forall (P_1, P_2) \in E \implies (P_1, P_2) \in E_P.$$

The number of connected components in $G_I[PS]$ is less than or equal to the cardinality of the set $PS$, such that all packages are reachable from the primary ones. Figure 1b shows the primary package subgraph for the VMI semantic graph in Figure 1a with Mariadb and Tomcat8 as primary packages.

A *base image subgraph* $G_I[BI]$ is the induced connected subgraph containing the base image and all dependency packages reachable from it:

$$G_I[BI] = (V_B, E_B), BI \in V_B \wedge V_B \subset V_I \setminus PS \wedge E_B \in E \wedge$$
$$\forall (P_1, P_2) \in E \implies (P_1, P_2) \in E_B.$$

The base image subgraph does not include primary packages. Figure 1c shows the base image subgraph for the VMI semantic graph in Figure 1a with Debian as base image.

### F. VMI semantic similarity

Let us consider two VMIs $I_1$ and $I_2$ with two semantic graphs $G_1(V_1, E_1)$ and $G_2(V_2, E_2)$. In this section, we define the similarity between two VMIs as a number in the $[0, 1]$ interval representing how similar the attributes of each vertex in $G_1$ and $G_2$ are to each other.

Initially, we define the *base image similarity* between two base images $BI_1$ and $BI_2$ as a binary digit with the value of 1 if the two images have the same attribute values and with the value of 0 otherwise:

$$sim_{BI}(BI_1, BI_2) = \begin{cases} 1, & attrs(BI_1) = attrs(BI_2), \\ 0, & attrs(BI_1) \neq attrs(BI_2). \end{cases}$$

Similarly, we define the *primary or dependency package similarity* between two packages $P_1$ and $P_2$ as 1 if they have the same name, version and architecture, also represented as binary values:

$$sim_{pkg}(P_1, P_2) = \begin{cases} 1, & pkg(P_1) = pkg(P_2), \\ 0, & pkg(P_1) \neq pkg(P_2); \end{cases}$$

$$sim_{ver}(P_1, P_2) = \begin{cases} 1, & ver(P_1) = ver(P_2), \\ 0, & ver(P_1) \neq ver(P_2); \end{cases}$$

$$sim_{arch}(P_1, P_2) = \begin{cases} 1, & arch(P_1) = arch(P_2) \vee \\ & arch(P_1) = \text{"all"} \vee \\ & arch(P_2) = \text{"all"}; \\ 0, & \text{otherwise}; \end{cases}$$

$$sim_P(P_1, P_2) = \prod_{\forall attr \in \{pkg, ver, arch\}} sim_{attr}(P_1, P_2).$$

An architecture attribute of `all` means that the package is portable and available on base images with any architecture.

Further, we compute the *size similarity* between two packages $P_1$ and $P_2$ as the ratio between the maximum size of the two packages divided by maximum size of all packages from both VMIs. Precisely, the size denotes the amount of disk space consumed by software package within a VMI, including any software package updates. Hence, we take the maximum value for sizes between the two software packages normalized over the union of all package sizes, which allows our model to compute a weighted composition of content and semantic similarity [5] of matched package within two VMIs:

$$sim_{size}(P_1, P_2) = \frac{\max\{size(P_1), size(P_2)\}}{\max\limits_{\forall P \in V_1 \cup V_2}\{size(P)\}}.$$

Finally, we model the *VMI semantic similarity* based on the Jaccard index [9], also known as intersection over union, which computes by what percentage a VMI semantic graph $G_1$ is similar to $G_2$. Hence, we formulate it as a product of the similarity between base images and the matched software packages with normalized package size in the numerator, and the union of all packages in both VMIs in the denominator:

$$Sim_G(G_1, G_2) = sim_{BI}(BI_1, BI_2) \cdot \frac{\sum\limits_{\forall (P_1, P_2) \in V_1 \times V_2} sim_{size}(P_1, P_2) \cdot sim_P(P_1, P_2)}{\sum\limits_{\forall (P_1, P_2) \in V_1 \times V_2} sim_{size}(P_1, P_2)}.$$

### G. Semantic compatibility

We define the *semantic compatibility* between a base image subgraph $G_I(BI) = (V_{BI}, E_{BI})$ and a primary package subgraph $G_I(PS) = (V_{PS}, E_{PS})$ as the product of the similarity values of their packages with a homonym $pkg$ attribute:

$$comp(G_I[BI], G_I[PS]) = \prod_{\substack{\forall (P_1, P_2) \in V_{BI} \times V_{PS} \\ \wedge pkg(P_1) = pkg(P_2)}} sim_P(P_1, P_2).$$

If the semantic compatibility has a value of 1, the primary packages can be installed and used together with the base image. Otherwise they are incompatible.

### H. VMI master graph

A *VMI master graph* $G_M[T, D, V, A]$ is a graph representation of all VMIs with same type, distribution, version and architecture base image attributes $(T, D, V, A)$ stored in the repository. The master graph contains one single base image subgraph semantically compatible to all primary package subgraphs of VMIs represented in the master graph. The purpose of the VMI master graph is to reduce the similarity computation overhead between multiple VMI semantic graphs with one single master graph similarity comparison. We therefore model the master graph as the union of one base image and one or more primary package subgraphs originating from VMIs with the same base image attributes:

$$G_M[T, D, V, A] = \bigcup_{\substack{\forall I \in Repo \wedge \\ attrs(BI) = (T, D, V, A) \wedge \\ comp(G_I[BI], G_I[PS]) = 1}} (G_I[BI] \cup G_I[PS]).$$

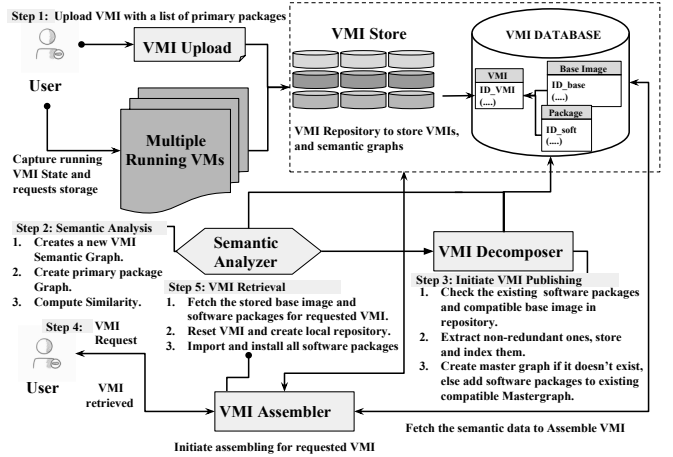

Fig. 2: Expelliarmus architecture and VMI management.

## IV. SEMANTIC-CENTRIC VMI MANAGEMENT

This section describes the architectural design of Expelliarmus, including the VMI publishing and retrieval algorithms.

### A. Architecture overview

Figure 2 describes the architecture of Expelliarmus through a use-case in which multiple users initially upload a VMI for storage in a proprietary Cloud image repository. Afterwards, they download and instantiate the VMI multiple times at various Cloud locations. Every time the users update a VMI and store it in the image repository, they introduce a considerable amount of redundancy and other management costs in terms of VMI publishing and retrieval, hindering the elastic provisioning and deployment process. At the same time, different VMIs with varying software packages uploaded by different users could also be composed of semantically similar base images and software packages. Expelliarmus semantically decomposes VMIs in reusable fragments so that similar software packages and base images within different VMIs are stored only once with reduced redundancy, publishing and retrieval overheads.

The publishing, storage and retrieval of a VMI in Expelliarmus takes place according to the following steps (see Figure 2):

1) The user uploads a VMI and a list of primary packages for storage in the VMI repository;
2) The semantic analyzer creates a VMI semantic graph and computes its semantic similarity with other VMIs;
3) To publish the VMI, the decomposer splits a VMI into a base image and multiple software packages exploiting the semantic similarity with other stored VMIs, such that only non-redundant packages and base images are stored again;
4) The user requests the retrieval of a VMI;
5) The VMI assembler assembles the VMI according to the user request and delivers it.

### B. VMI semantic analyzer

The semantic analyzer takes the VMI and the primary package list as input and constructs the semantic graph, following the model defined in Section III-B. This enables an automated approach for optimizing and accessing semantic similarity of monolithic VMIs without detailed content analysis, instead caching a subset of VMI semantic data in the form of a graph.

The semantic analyzer creates a graph $G_I$ for every uploaded VMI $I$ along with a subgraph representation of the corresponding primary packages and base image. Afterwards, the semantic analyzer compares the newly uploaded VMI with the appropriate master graph $G_M$ having the same type, distribution, version and architecture attributes, previously stored in the repository according to the semantic similarity defined in Section III-F. Every VMI master graph is specific to a characteristic base image with one or more semantically compatible primary package subgraphs. If no such master graph exists in the repository, the semantic analyzer forwards the VMI to the VMI decomposer in either case.

### C. VMI decomposer

The VMI decomposer splits a VMI into a base image and different software packages, exploiting semantic similarity such that only non-redundant software packages and base images are stored. To achieve this, decomposer employs two algorithms: VMI publishing (Algorithm 1) and base image selection (Algorithm 2).

*1) VMI publishing algorithm:* Algorithm 1 outlines the step-wise VMI publishing process. The algorithm takes as input a VMI $I$, its semantic graph $G_I$, a VMI repository, and list of primary packages $PS$. Initially, the algorithm extracts the VMI's primary package subgraph $G_I[PS]$ in line 1. Afterwards, it iterates each primary package in the subgraph, checks if it exists in the repository with same semantic attributes (lines 2 – 5) and, if it does not exist, stores it (line 4) in the repository. After checking all primary subgraph packages, the algorithms stores the user data in line 6. Next, line 8 removes the primary packages from the VMI, including the user data and the dependency packages not used by any software package still within the VMI (lines 10 – 11). At this point, the VMI contains only the base image $BI$ (line 12) with all its required software packages already stored in the repository. To prevent redundant storage of the same base image, a base image selection algorithm (see Algorithm 2 in Section IV-C2) called at line 14 returns a similar base image together with a list of base images stored in the repository that are no longer required. If Algorithm 2 returns the current base image, we update the repository along with the corresponding new master graph in lines 15 – 17. However, if the Algorithm 2 returns another already stored base image, line 19 retrieves its master graph from the repository and updates the retrieved master graph with the primary package subgraph $G_I[BI]$ in line 21. Algorithm 2 also returns a list of base images that can be replaced with the selected base image. Line 22 – 28 iterates over this list and line 23 retrieves the master graph for each replaceable base image from the repository. Each retrieved master graph corresponding to a base image in the list is a union of a base image subgraph and several primary package subgraphs. Lines 24 – 26 iterate over the primary packages in each master graph and update the master graph of the selected base image $G_M$ with the extracted primary package subgraph (line 25). Line 27 removes the obsolete base images and line 29 updates the master graph in the repository.

**Algorithm 1:** VMI publishing algorithm.

**Input** : $I = (BI, PS, DS, Data)$: VMI; $G_I$: semantic graph of VMI $I$; $PS$: primary package set; $repo$: VMI repository

1  $G_I[PS] = (V_P, E_P) \leftarrow$ extractSubGraph$(G_I)$
2  **forall** $P \in V_P$ **do**
3      **if** $\neg$ exists$(P, repo)$ **then**
4          store$(P, repo)$
5  **end**
6  store$(Data, repo)$
7  **forall** $P \in PS$ **do**
8      remove$(P, I)$
9  **end**
10 removeUnusedDependencies$(I)$
11 remove$(Data, I)$
12 $BI \leftarrow I$
13 $G_I[BI] \leftarrow$ createSubGraph$(BI)$
14 $(base, list) \leftarrow$ selectBaseImage$(BI, G_I[BI], G_I[PS], repo)$
15 **if** $base = BI$ **then**
16     $G_M \leftarrow$ createMasterGraph$(G_I[BI])$
17     store$(BI, repo)$;
18 **else**
19     $G_M \leftarrow$ getMasterGraph$(base, repo)$
20 **end**
21 $G_M \leftarrow G_M \cup G_I[PS]$
22 **forall** $b \in list$ **do**
23     $G_{Mb} \leftarrow$ getMasterGraph$(b, repo)$
24     **forall** $P \in G_{Mb}$ **do**
25         $G_M \leftarrow G_M \cup$ extractSubGraph$(G_{Mb}, P)$
26     **end**
27     remove$(b, repo)$
28 **end**
29 update$(G_M, repo)$

*2) Base image selection algorithm:* As a part of the VMI publishing, Algorithm 2 returns an appropriate base image and a replace list of previously stored base images no longer needed. The selected base image is semantically compatible with the primary packages corresponding to the base images in the replace list. The algorithm takes as input a base image $BI$, the primary package subgraph $G_I[PS]$ of an image $I$ and a VMI repository $repo$. Initially, line 1 initializes a triplet list with the base image of a VMI $I$, the base image subgraph and the primary package subgraph. Afterwards, line 3 retrieves the list of all base images stored in the repository. Lines 4 – 12 iterate over the list of stored base images, and gets the corresponding base image subgraph and master graph in lines 5 and 6. Next, the algorithm checks the semantic similarity between the base image $BI$ and the stored base images in line 7. If the semantic similarity between the base images exist, line 9 extracts each primary package subgraph from the master graph, while line 10 adds into a triplet list the stored base image, its base image subgraph, and the primary package subgraphs. Afterwards, lines 13 – 26 iterate over all base images in this triplet list. For each current base image, the algorithm adds first into the replace list all other base images $BI_j$ that are not identical but similar $B_i$, and contain semantically compatible (see Equation III-G) primary packages (see lines 15 – 18, meaning that the current base image $BI_i$ can replace all the base images in its replace list). If the replace list is not empty (line 20), line 23 computes the total size of all packages of the base image $B_i$ in the replace list. Finally, line 25 adds the base image, the replace list, and the total size of its primary packages into a new quadruple list. The fourth boolean component (i.e. $BI_i = BI$) of the quadruple indicates whether this base image is new or already existed in the repository. Once this procedure completes for all base images in the triplet list, line 27 sorts the generated quadruples list based on three criteria: the replace list size

**Algorithm 2:** Base image selection algorithm.

**Input** : $BI$: remaining base image after decomposition; $G_I[BI]$: base image subgraph; $G_I[PS]$: primary package subgraph; $repo$: VMI repository

1. $list3 \leftarrow [(BI, G_I[BI], G_I[PS])]$
2. $list4 \leftarrow \emptyset$
3. $baseList \leftarrow$ getBaseImageList($repo$)
4. **forall** $b \in baseList$ **do**
5.     $G_I[b] \leftarrow$ getSubGraph($b, repo$)
6.     $G_{Mb} \leftarrow$ getMasterGraph($b, repo$);
7.     **if** $sim_{BI}(BI, b) = 1$ **then**
8.         **forall** $P \in G_{Mb}$ **do**
9.             $G_I[P] \leftarrow$ extractSubGraph($G_{Mb}, P$)
10.             $list3 \leftarrow list3 \cup \{(b, G_I[b], G_I[P])\}$
11.         **end**
12. **end**
13. **forall** $i \in list3$ **do**
14.     $(BI_i, G_I[BI_i], G_I[PS_i]) \leftarrow i$
15.     **forall** $j \neq i \in list3$ **do**
16.         $(BI_j, G_I[BI_j], G_I[PS_j]) \leftarrow i$
17.         **if** $BI_i \neq BI_j \wedge comp(G_I[BI_i], G_I[PS_j]) = 1$ **then**
18.             $replaceList \leftarrow replaceList \cup \{BI_j\}$
19.     **end**
20.     **if** $replaceList \neq \emptyset$ **then**
21.         $size \leftarrow 0$
22.         **forall** $P \in G_I[BI_i]$ **do**
23.             $size \leftarrow size + size(P)$;
24.         **end**
25.         $list4 \leftarrow list4 \cup \{(BI_i, replaceList, size, BI = BI_i)\}$
26. **end**
27. $list4 \leftarrow$ sort($list4$)
28. **forall** $i \in list4$ **do**
29.     $(BI_i, replaceList, \_, \_) \leftarrow i$
30.     **if** $BI_i = BI \vee BI \in replaceList$ **then**
31.         **return** $(BI_i, replaceList)$
32. **end**
33. **return** $(BI, \emptyset)$

---

**Algorithm 3:** VMI retrieval algorithm.

**Input** : $I = (BI, PS, \_, Data)$: VMI; $repo$: VMI repository

1. $(G_I[BI], G_I[PS]) \leftarrow$ getSubGraph($I, repo$)
2. **if** $G_I[BI] \neq$ NULL $\wedge G_I[PS] \neq$ NULL $\wedge comp(G_I[BI], G_I[PS]) = 1$ **then**
3.     $BI \leftarrow$ getBaseImage($id, G_I[BI], repo$)
4.     resetVMI($BI$)
5.     import($Data, I$)
6.     **forall** $P \in G_I[PS]$ **do**
7.         **if** $P \notin G_I[BI]$ **then**
8.             $PS \leftarrow PS \cup P$
9.         **end**
10.     **end**
11.     **forall** $P \in PS$ **do**
12.         install($P, I$);
13.     **end**
14. **end**
15. **return** $I$

---

(i.e. the more replaced base images, the better), the total size of its primary packages (the smaller, the better), and existence of a similar base image in the repository (i.e. no unnecessary storage). Lines 29 – 32 iterate over the sorted quadruples list, and extract the base image and its replace list in line 29. Finally, it checks the first quadruple that either specifies the base image $BI$ or exists in the replace list in line 30 and returns it in line 31. If no quadruple exists with the base image $BI$, line 33 returns it with an empty replace list.

### D. VMI assembler

Expelliarmus enables VMI assembly either with identical or with differing functionality, provided that the requested software package exists in the repository. To achieve this, the VMI assembler employs a VMI retrieval algorithm (Algorithm 3) that processes the VMI retrieval requests for deployment.

*1) VMI retrieval algorithm:* Algorithm 3 represents the stepwise VMI retrieval procedure using two input parameters: a (nonexistent) VMI $I$ identified by its base image $BI$ and primary package set $PS$, and a VMI repository $repo$. Initially, line 1 obtains the base image and primary package subgraphs from the repository. If they exist and are compatible (line 2), line 3 retrieves the base image $BI$ from the repository and resets it to an initial state in line 4. Line 5 imports the user data into the VMI $I$. Lines 6 – 10 iterate over each vertex in the primary package subgraph and check their existence in the base image subgraph in line 7. If the vertex does not exist, line 8 adds the vertex into the primary package set $PS$. Finally, the VMI's guest OS package manager installs the primary packages in line 12 by importing the required software packages (including primary and dependency packages), if necessary. Line 15 returns the assembled VMI.

## V. IMPLEMENTATION

We implemented Expelliarmus in Python and publicly released the complete source code including the reproducibility artifact[1] of our implementation and experimental validation in GitHub[2]. In the following, we briefly discuss the implementation details of Expelliarmus with respect to VMI manipulation, semantic graph creation, publish and retrieval, currently limited to Linux VMIs only.

*1) VMI access:* Expelliarmus uses the `libguestfs`[3] library to access, manipulate and modify VMIs. Apart from performing modifications to a VMI file system, `libguestfs` provides access to the guest OS through its virtual appliance without instantiating the entire VMI. To achieve this, `libguestfs` configures and launches a `guestfs` handle that provides an interface to access VMIs.

*2) VMI graph representation:* We represent VMIs according to semantic principles described in Section III-B and store them in graph data structure using the Python-based `networkx`[4] module, implemented depending upon the suitable package management of the guest operating system (e.g. `APT` or `DNF`). We execute the package management commands through `libguestfs` on the VMI guest OS to fetch the required semantic information (e.g. architecture, version) about the base image, installed software packages and dependency packages, associated to graph vertices and edges.

*3) VMI publishing:* comprises decomposition process, accomplished by recreating the binary package (e.g. `.deb` distribution files) for the required software packages and utilizing the `libguestfs` calls to export them to the VMI repository. Furthermore, we remove the specific package binaries, configurations and the dependency packages no longer required in the VMI, followed by cleaning up the cached repository files. Finally, we employ the base image selection algorithm to select the appropriate base image for storage.

---

[1] https://github.com/ExpelliarmusSuperComp/Expelliarmus/blob/master/ReproducibilityArtifact.pdf
[2] https://github.com/ExpelliarmusSuperComp/Expelliarmus
[3] http://libguestfs.org
[4] https://networkx.github.io/

*4) VMI retrieval:* comprises the assembly procedure, achieved by first resetting the base image using the `virt-sysprep` tool (part of `libguestfs`), followed by importing software packages and specific user data into the VMI. Furthermore, we scan the imported software packages that create the meta-data for each imported package readable by the package manager. Afterwards, we add a custom repository configuration file (i.e. pointer to the software packages in the local repository) that enables the VMI's guest OS package management to install packages from the local repository instead of the online ones. Finally, we remove the local temporary repository including the custom configuration file and restore the default repository configuration files (i.e. pointer to the online package management repository).

## VI. EXPERIMENTAL RESULTS

### A. Experimental setup

We implemented Expelliarmus on a quad-core machine with Ubuntu 16.04 (`x86_64`) OS architecture and external SSD disk, with storage capacity of $1\,\mathrm{TB}$ acting as a VMI repository. We further used the SQLite[5] database engine, suitable for managing VMI meta-data due to its self-contained, serverless, and zero-configuration characteristics. In principle, our system is capable of running on any Linux-based OS with support for `libguestfs`, `qemu` and `SQLite` software tools.

In the lack of any public VMI management benchmark, we evaluate our approach using a synthetic VMI set based on the Ubuntu Linux distribution with software packages recognized by package management tools. We will address the management of VMIs composed of software packages recognized by non-package management tools (e.g. `pip`, `snap`) or installed through compiled source code in future work. We create each VMI using `virt-builder`[6], an efficient tool for building a variety of images for local and Cloud use. The minimal script to create an image in our experiments is available in the GitHub code repository[7]. For a fair comparison, the evaluation set includes the four VMIs used in two previous studies [1], [15], namely Mini, Base, Desktop, and IDE, in the same configuration. To provide a representative set of Cloud images for the evaluation, the VMI set also includes 15 images with a similar software stack as provided at the Amazon Web Services portal[8] for free and enterprise use:

1) *Mini* image with non-desktop minimal installation of Ubuntu Linux distribution;
2) *Base* image with LAMP software stack;
3) *Desktop* image with X Windows and desktop productivity tools including LAMP software stack, FTP/NFS and email servers;
4) *IDE image* with integrated development environment including Eclipse, JDK, and Python;
5) *AWS infrastructure software* images with application servers (e.g. LAPP, LEMP, Jenkins, Tomcat, MongoDB, CouchDB, Django, RabbitMQ, Redis, Cassandra, PostgreSQL, Elastic stack), and project management tools (i.e. Redmine, ownCloud).

[5]https://www.sqlite.org/index.html
[6]http://libguestfs.org/virt-builder.1.html
[7]https://github.com/ExpelliarmusSuperComp/Expelliarmus
[8]https://aws.amazon.com/marketplace

TABLE II: Experimental VMI characteristics.

| VMI number | VMI name | Mounted size [GB] | Number of files | Similarity $[Sim_G]$ | Publishing time [s] | Retrieval time [s] |
|---|---|---|---|---|---|---|
| 1 | Mini | 1.913 | 75749 | 0 | 39.52 | 24.64 |
| 2 | Redis | 1.914 | 75796 | 0.97 | 10.28 | 22.05 |
| 3 | PostgreSql | 1.963 | 77497 | 0.59 | 39.699 | 33.91 |
| 4 | Django | 1.969 | 79751 | 0.71 | 18.916 | 27.30 |
| 5 | RabbitMQ | 1.956 | 77596 | 0.56 | 25.620 | 33.87 |
| 6 | Base | 1.986 | 78471 | 0.89 | 42.236 | 47.17 |
| 7 | CouchDB | 1.965 | 77725 | 0.70 | 37.99 | 42.58 |
| 8 | Cassandra | 2.531 | 79740 | 0.71 | 42.58 | 35.66 |
| 9 | Tomcat | 2.049 | 76356 | 0.37 | 60.65 | 36.37 |
| 10 | Lapp | 2.107 | 77816 | 0.53 | 56.71 | 61.79 |
| 11 | Lemp | 2.112 | 77360 | 0.97 | 25.093 | 57.11 |
| 12 | MongoDb | 2.110 | 75820 | 0.15 | 90.465 | 29.33 |
| 13 | Own Cloud | 2.378 | 90667 | 0.76 | 80.942 | 100.43 |
| 14 | Desktop | 2.233 | 90338 | 0.50 | 201.721 | 102.34 |
| 15 | Apache Solr | 2.338 | 79161 | 0.84 | 71.555 | 92.57 |
| 16 | IDE | 2.727 | 81200 | 0.52 | 135.333 | 63.62 |
| 17 | Jenkins | 2.515 | 79695 | 0.87 | 63.504 | 81.24 |
| 18 | Redmine | 2.363 | 95309 | 0.79 | 112.908 | 97.08 |
| 19 | Elastic Stack | 2.671 | 103719 | 0.64 | 166.001 | 99.91 |

Table II lists the characteristics of the VMIs including their mounted disk use, number of files in their file system, semantic similarity, as well as the publishing and retrieval times to and from the repository. We average the evaluation results across five trials, as their variance is relatively small in all experiments. We assume that the repository is initially empty and the VMIs are randomly uploaded to the repository for the first execution. To maintain the uniformity of evaluation over five trials, we perform the remaining four executions in the same sequence listed in Table II.

### B. VMI repository optimization

To evaluate the storage optimization achieved by Expelliarmus, we consider three separate scenarios. Initially, we show the cumulative repository size by adding four VMIs from previous studies [1], [15] namely Mini, Base, Desktop, and IDE. In the second scenario, we upload the VMIs listed in Table II to show the growth of repository at larger scale. Finally and similar to previous studies [1], [15], the third scenario evaluates the storage performance of the repository by adding 40 IDE images obtained by successive builds. For each scenario, we compare the Expelliarmus storage efficiency with the following VMI encoding schemes:

- *Qcow2* format with no compression[9];
- *Qcow2 + Gzip* compressed format[10];
- *Mirage* MIF [1] format that uses a manifest file to manage the VMI content descriptors and store the image contents in a global data store;
- *Hemera* [15] format that uses a file system and database oriented hybrid approach for managing image content.

Figure 3a shows the cumulative repository growth for the first scenario of adding four successive images (Mini, Base, Desktop, and IDE). On a repository that stores only four images with a cumulative size of $8.85\,\mathrm{GB}$ in Qcow2 format, Expelliarmus performs better requiring only $2.3\,\mathrm{GB}$, as compared to $3.2\,\mathrm{GB}$ for images compressed with Gzip, and $3.4\,\mathrm{GB}$ for Mirage and Hemera systems. Figure 3b estimates the performance of these systems for adding 19 VMIs, successively listed in Table II. For a cumulative repository size of 41.81 GB in the Qcow2 format, Expelliarmus requires 2.75 GB, while Mirage and Hemera perform again similarly requiring 8.81 GB. In this scenario, the storage cost for Qcow2 images compressed with Gzip encoding scheme is worse requiring 15 GB. An important observation deduced from Figure 3a and Figure 3b is the improved performance of Mirage, Hemera and Expelliarmus with increasing number of VMIs in the repository over Gzip-based scheme. The performance of Mirage and Hemera owes to file level deduplciation that stores common files from the same and different VMIs only once. Expelliarmus instead, not only relies on deduplicating similar software packages, but also optimizes the VMI storage by removing software packages not a dependency for the primary ones, an opportunity not captured by Mirage and Hemera systems. Moreover, Expelliarmus's base image selection reduces the storage by selecting one from a pool of semantically similar base images, while Mirage and Hemera store additional non-redundant content of other base images. Evidently, the base image is a major contributor to the higher repository size.

[9]https://people.gnome.org/~markmc/qcow-image-format.html
[10]https://www.gzip.org/

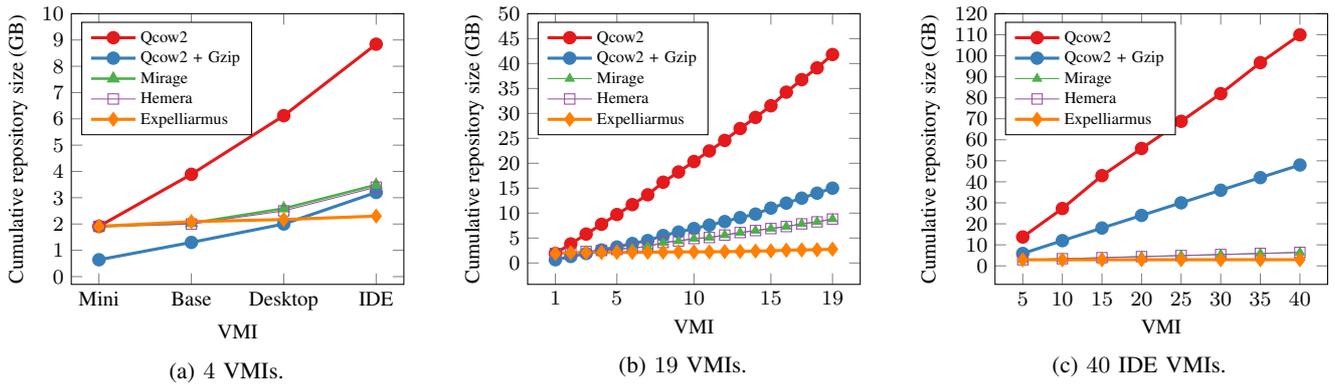

Fig. 3: Repository size growth with different numbers of successively stored VMIs.

The advantage of Expelliarmus is better represented by the third scenario shown in Figure 3c, including 40 IDE-similar VMIs composed of the same base image and software packages. For a cumulative repository size of 109.92 GB, Expelliarmus requires 2.94 GB, while Mirage and Hemera-based storage require 6.4 GB. The storage cost for the Gzip compressed scheme is even higher, requiring 48 GB. In this scenario, Expelliarmus performs 16 times better than Gzip, and 2.2 times better than Mirage and Hemera, which in turn perform 7.5 times better than Gzip.

## C. VMI publishing and retrieval

We further evaluate the performance of publishing and retrieving different VMIs to and from the repository. The time of publishing a VMI in Expelliarmus represents the decomposition time, comprising time to create `guestfs` handle for VMI access, export semantically non-redundant software packages, remove the unused software packages, and select the compatible base image. On the contrary, the time to retrieve a VMI reflects the assembly time comprising the time to create a `guestfs` handle, copy the appropriate base image to the local repository, reset the VMI, and import software packages. We evaluate VMI publishing time for two scenarios. The first scenario represents the sequential upload of four VMIs used in previous studies [1], [15] (Mini, Base, Desktop, and IDE), while the second scenario evaluates all VMIs listed in Table II (including the ones from the first scenario).

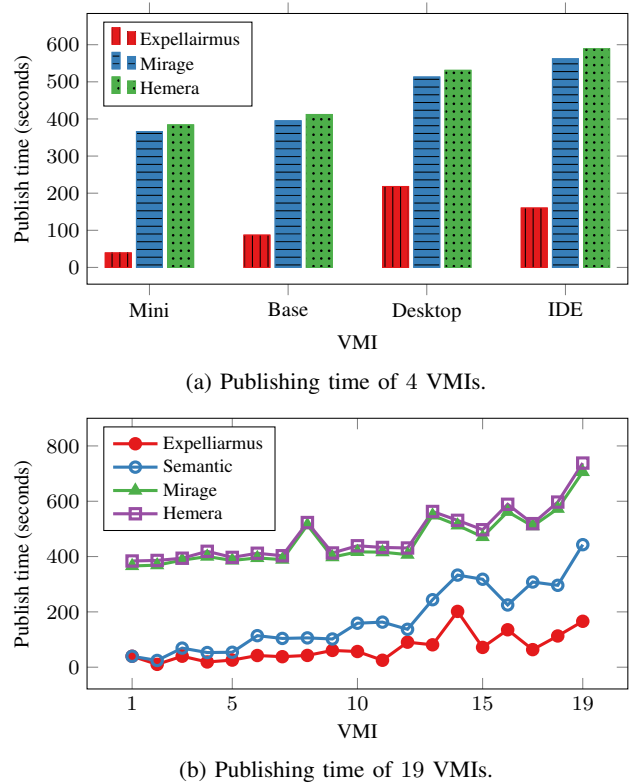

(a) Publishing time of 4 VMIs.

(b) Publishing time of 19 VMIs.

Fig. 4: VMI publishing time analysis.

Figure 4a shows the VMI publishing time for the first scenario. The Expelliarmus optimizes not only the storage cost as previously discussed, but also publishes VMI faster compared to both Mirage and Hemera. The publishing time of a VMI in Expelliarmus depends not only on the mounted VMI size, but also on the software packages installation size. The installation size is the amount of space required by the software package to be installed on a disk, which is always larger than

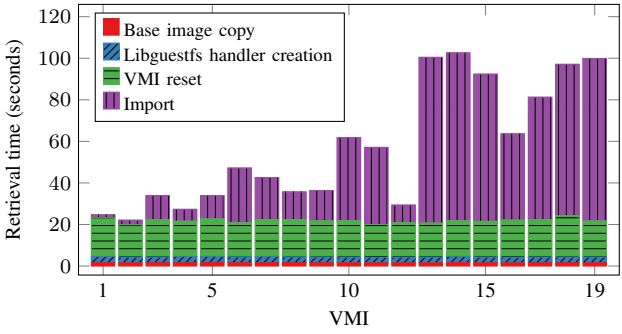

(a) Retrieval time for Expelliarmus.

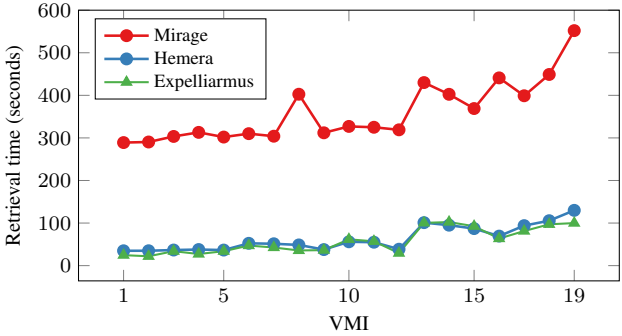

(b) Retrieval time comparison for VMIs listed in Table II.

Fig. 5: Retrieval time analysis of VMIs.

the size of a software packaged in the .deb or .rpm format. The different software packages with varying installation sizes largely affect the time to create a binary software package (e.g. .deb) resulting in a higher export time of the same to the repository. The total installation size of the exported software packages for the Desktop VMI is the largest, and hence requires more time to publish in Expelliarmus compared to other images. In contrast, Mirage and Hemera require more time for the IDE VMI, as the publishing time is proportional to the mounted size and the file sizes within a VMI.

Another reason for the better VMI publishing time compared to the file system-based approaches is due to lower deduplication overheads. Mirage and Hemera require matching content over thousands of files incurring time penalties in the range of seconds to few minutes. In contrast, Expelliarmus relies on VMI semantic graphs for similarity computation and semantic clustering of similar VMIs into a master graph, which allows the comparison of new VMIs with a single master graph instead of multiple VMIs. In Expelliarmus, the similarity computation incurs time penalties in the order of less than $100\,\text{ms}$ for each VMI. This eradicates a large share of VMI publishing overhead with low similarity computation cost compared to Mirage and Hemera.

Figure 4b shows the publishing time over a repository with 19 VMIs, successively added as listed in Table II, representing the second scenario. In this case, we additionally use for comparison a variant of Expelliarmus called *semantic decomposition* that exports all the required software packages without taking semantic similarity into account. While for the Mirage and Hemera systems the Elastic Stack VMI requires the highest publishing time due to its mounted size and large number of files (100 thousand), the Desktop VMI had the longest publishing time in Expelliarmus followed by Elastic Stack. Interestingly, the total installation size of packages to export for both Desktop and Elastic Stack VMIs is nearly equal, yet the former takes more time to publish compared to the latter. The reason is that the system requires to export 126 software packages for the Desktop VMI, compared to only three packages for Elastic Stack. For semantic decomposition, the longest publishing time is for Elastic Stack VMI, with Expelliarmus performing better compared to its variant (expected, as it only exports the software packages that do not exist in the repository). As previously discussed, the similarity computation overhead in Expelliarmus is minimal, hence the VMI publishing time is majorly dependent on the export of software packages. However, with upload of more VMIs to the repository, Expelliarmus require less software packages to export with lower publishing time contrary to its variant.

To evaluate the VMI retrieval, we consider only one scenario of a repository with 19 images listed in Table II. Figure 5a shows the VMI retrieval in Expelliarmus as a composition of four operations: copying the base image from the repository, creating the `guestfs` handler, resetting the VMI, and finally importing the required software packages into a VMI. The first three operations share nearly equal time for retrieving different VMIs, while the import time differs invariably. Similar to VMI publishing, the VMI retrieval in Expelliarmus depends on the installation size of the imported software packages, which is highest in case of Desktop VMI.

Figure 5b compares the VMI retrieval time, which is fastest for Hemera and Expelliarmus than Mirage. Expelliarmus's better performance is due to selective package retrieval (including only primary and corresponding dependency packages) imported into a VMI that significantly reduces the total size retrieved from the repository. Mirage's VMI retrieval is worse for two reasons: (1) it retrieves more data by reading many files instead of reading linearly through one file, and (2) it is inefficient in reading small files (below $1\,\text{MB}$) from file system-based repository. Hemera improves this overhead using a hybrid approach that stores large files in the repository and small sized files in the database, which optimizes VMI retrieval as the database handles small files much faster than the file system. Although, Hemera and Expelliarmus perform nearly equal for most VMIs, the retrieval time of Elastic Stack VMI is slightly different in both cases. While Expelliarmus retrieval takes $99.9\,\text{s}$, Hemera needs $129.8\,\text{s}$ mostly due to retrieving a large number of files (more than 100 thousand).

## VII. CONCLUSION

We introduced Expelliarmus, a new VMI management system with a semantic-centric design for VMI storage with optimized VMI publish and retrieval. Different from the existing VMI management systems that ignore VMI semantics, Expelliarmus incorporates three features. First, it represents VMIs as structured semantic graphs, efficiently expressing the functional requirements between the base image and the different software packages. Such an approach allows clustering of multiple VMIs into a single master graph and thus expedites similarity computation. Second, Expelliarmus enables a semantic aware VMI decomposition and base image selection to extract and subsequently store only non-redundant base image and software packages. Third, Expelliarmus is capable of performing VMI assembly on the fly, either by fetching initially uploaded software packages or by selecting compatible and semantically similar ones already existing in the repository. We evaluated Expelliarmus over representative set of synthetic Cloud VMIs on a real testbed. Results show that semantic-centric management of VMIs is able to reduce the repository size by $2.2 - 16$ times compared to three related systems, and significantly improves the VMI publish and retrieval performance. Currently, Expelliarmus supports Linux VMIs, while managing Windows VMIs is part of our future work. We also plan in the future to extend Expelliarmus to support automated containerization of a VMI with multiple container service functionality.